\documentclass[pre, aps, showpacs]{revtex4}
\usepackage[dvips]{graphicx}
\usepackage{mathrsfs}
\usepackage{latexsym}
\begin{document} 
%
%%%%%%%%%%%%%%%%%%%%%%%%%%%%%%%%%%%%%%%%%%%%%%%%%%%%%%%%%%%%%%%%%%%%%%%%%%%% 
%
\title{Protein-Mediated DNA~Loops: Effects of Protein Bridge Size and Kinks} 
\author{Nicolas~\textsc{Douarche} and Simona~\textsc{Cocco}} 
\affiliation{CNRS -- Laboratoire de Physique Statistique de l'ENS, 
24~rue Lhomond, 75005~Paris, France} 
\begin{abstract}
This paper focuses on the probability that a portion of~DNA closes on itself 
through thermal fluctuations. We investigate the dependence of this probability 
upon the size~$r$ of a protein bridge and/or the presence of a kink at half 
DNA~length. The~DNA is modeled by the Worm-Like Chain model, and the probability 
of loop formation is calculated in two ways: exact numerical evaluation of the 
constrained path integral and the extension of the \textsc{Shimada} 
and~\textsc{Yamakawa} saddle point approximation. For example, we find that the 
looping free energy of a 100~base~pairs~DNA decreases from 24~$k_\mathrm{B}T$ to 
13~$k_\mathrm{B}T$ when the loop is closed by a protein of $r=10$~nm length. 
It further decreases to 5~$k_\mathrm{B}T$ when the loop has a kink of 
$120^\circ$ at half-length. 
\end{abstract}
\maketitle
%
%%%%%%%%%%%%%%%%%%%%%%%%%%%%%%%%%%%%%%%%%%%%%%%%%%%%%%%%%%%%%%%%%%%%%%%%%%%%
%
\section{Introduction: DNA Loops in Gene Transcription Regulation} 
Gene expression is regulated by a wide variety of mechanisms. These activation 
as well as repression phenomena may occur at every expression steps~(translation, 
transcription, etc) and do involve interactions between several biological 
molecules~(DNA, RNAs, proteins, etc). For instance proteins bound on specific 
DNA~sequences may turn~on/off genes transcription by interacting with each other. 
By bringing closer those proteins DNA~looping can ease their 
interactions~\cite{rippe95, focus, finzi, lia}. Such looping events may be 
observed over a wide range of lengths spreading from hundreds to thousands base 
pairs~(bp). The looping probability has been firstly measured by the cyclization 
of DNA~segments in solution with cohesive ends. Once the loop is formed proteins 
called ligases stabilize it~\cite{shore, du}. It is then possible to count the 
circular~DNAs with respect to the linear ones. The loop formation mediated by 
proteins have also been experimentally studied. Two examples are the loops 
formed by the~LacR or the~GalR transcriptional repressors~\cite{brenowitz}. 
Two units of such proteins bind at two specific positions along a same~DNA and 
associate to form a complex when the binding sites come in contact. The 
formation of such loops has been recently studied using micromanipulation 
experiments on a single DNA~molecule~\cite{finzi, lia}. The study of the~GalR 
mediated loop has shed light on the role of a third protein called~HU that 
sharply bends~(\emph{i.e.} kinks) the~DNA at half-length. \\
The DNA~loop probability depends mainly on its length and flexibility. 
Long~DNAs~(typically longer than 1500~bp) essentially behave as Gaussian 
Polymers~(GP)~\cite{yamakawa, kleinert}: the cyclization cost is mainly of an 
entropic nature~\cite{hanke}. On the contrary, for small lengths DNA~cyclization 
is difficult mainly because of the bending energy cost. The computation of the 
elastic energy for the Worm-Like Chain~(WLC) model~\cite{kratky, yamakawa, kleinert} 
can be analytically performed~\cite{shimada}; numerical methods have also been 
employed when electrostatic properties are included~\cite{balaeff}. At 
intermediate length scales~(from about 500~bp) elastic rigidity and entropic 
loss are both important. Several approximations have been developed to study 
this lengths range~\cite{focus, daniels, gobush, wilhelm}, among them the 
calculations of the fluctuations around the lowest bending energy configurations 
performed by~\textsc{Shimada} and~\textsc{Yamakawa}~\cite{shimada}. Numerical 
approaches have also been developed: Monte~Carlo~\cite{podtelezhnikov} and brownian 
dynamics based simulations~\cite{rippe95, merlitz, rippe00} as well as numerical 
calculations of the~WLC path integral under the closed ends constraint~\cite{yan03,yan04}. 
This last method allowed~\textsc{Yan}, \textsc{Kawamura}~and~\textsc{Marko} to 
study the elastic response of~DNA subject to permanent or thermally excited bendings 
caused by binding proteins~(such as~HU) or  inhomogeneities along the DNA~double 
helix~\cite{yan04,yan05}. All these studies do provide a better understanding of the 
underlying regulation phenomena despite their overall complexity. \\ 
In this paper we study two processes that turned out to be important in 
DNA~looping. Namely the size of the protein complex clamping the 
loop~\cite{rippe95, merlitz}, acting as a bridge between the two DNA~ends, and 
mechanisms implying DNA~stiffness loss which are taken into account in an 
effective way by kinking the~WLC at half-contour length~\cite{rippe95, merlitz, %
rippe00, sankararaman}. In section~\ref{s:defmeth} we define the model 
and the methods: we describe the numerical approach~(\S~\ref{secnum}) and the 
analytical Saddle Point Approximation~(SPA, \S~\ref{s:sp1}) to calculate the 
$r$-dependent closure factor and the looping free energy. In section~\ref{s:size} 
we compare the numerical and~SPA  results with previous experimental and 
theoretical results. In section~\ref{s:kink} we extend the numerical and~SPA 
approaches to a kinked loop; we discuss our results and we propose a simple 
formula that accounts for both the protein bridge and kink 
effects~(\S~\ref{ss:kink}). We conclude (section~\ref{s:conclusion}) by sketching 
how to include omitted DNA~properties which may also play an important role in 
its closure such as twist rigidity or electrostactic interactions. 
% 
%%%%%%%%%%%%%%%%%%%%%%%%%%%%%%%%%%%%%%%%%%%%%%%%%%%%%%%%%%%%%%%%%%%%%%%%%%%%%%%% 
%
\section{Definitions and Methods}\label{s:defmeth}
We use the well known Worm-Like Chain~(WLC) model~\cite{kratky, yamakawa, kleinert}. 
The DNA~polymer is described as an \emph{inextensible} continuous differentiable 
curve of contour length~$L$, with unit tangent 
vector~$\vec t(s)$~($0 \le s \le L$). The polymer is characterized by the 
persistence length~$A$ beyond which tangent vectors lose their alignment: 
$\langle \vec t(s) \cdot \vec t(s') \rangle = %
\exp{\! \left(-\vert s-s' \vert/A\right)}$. The energy of a configuration 
of the polymer stretched under an external force~$f \, \vec e_z$ reads 
\begin{equation} 
E \! \left[\vec t ; L, f\right] = \frac{1}{2} \frac{A}{\beta} \int_0^L 
\! \left[\frac{\mathrm{d}\vec t(s)}{\mathrm{d}s}\right]^2 \!\!\! \mathrm{d}s 
- f \int_0^L \!\! \vec e_z \cdot \vec t(s) \, \mathrm{d}s 
\end{equation} 
where we use $\beta=1/k_\mathrm{B}T$. No twist elasticity nor extensibility 
will be considered. The partition function is 
\begin{equation}\label{defz}
Z(L, f) = \int \!\! \mathscr{D}\vec t \; 
\exp{\! \left\{-\beta E \! \left[\vec t ; L, f\right] \right\}}. 
\end{equation}
Notice that summation over all initial and final tangent vectors orientations, 
$\vec t(0)$ and~$\vec t(L)$, is implicitly understood in this path integral. \\ 
In this paper we are interested in the formation of a loop in a DNA~molecule, 
and the Probability Density Functions~(PDFs) of end-to-end distances play an 
important role. The quantities under study are denoted by~$Q$,~$S$,~$P$ and~$J$ 
respectively and defined as follows: 
% 
%%%%%%%%%%%%%%%%%%
%
\begin{itemize}
\item the end-to-end extension~$\vec r = (x, y, z)$~PDF at zero force, 
\begin{equation}\label{Q} 
Q \! \left(\vec r , L\right) = \frac{1}{Z(L, f=0)} \, 
\int \!\! \mathscr{D}\vec t \,\,\, 
\delta \! \left[\int_0^L \!\! \vec t(s) \, \mathrm{d}s - \vec r \right] 
\exp{\! \left\{-\beta E \! \left[\vec t ; L, f=0\right]\right\}}. 
\end{equation} 
In the absence of force $Q$~depends on its argument~$\vec r$ through its 
modulus~$r = \vert \vec r \vert$ only, and we may introduce the radial~PDF 
\begin{equation}
S(r, L) = 4 \pi r^2 \; Q \! \left(r, L\right). 
\end{equation}
\item The $z$~extension~PDF reads 
\begin{equation}\label{P} 
P(z, L) = \int_{-L}^L \!\!\!\!\! \mathrm{d}x 
\int_{-L}^L \!\!\!\!\! \mathrm{d}y \,\, Q\left[(x, y, z), L\right]. 
\end{equation}
In the absence of external force, notice the~$x$ or~$y$ extensions~PDFs are 
given by~$P$ too. Interestingly, the radial and $z$~extensions~PDF are related 
to each other through the useful identity~\cite{samuel} 
\begin{equation}\label{relasp}
S(r, L) = -2r \; \frac{\mathrm{d}P}{\mathrm{d}z}(r, L). 
\end{equation}
\item The cyclization factor 
\begin{equation}
J(L) = Q(0, L) 
\end{equation}
defined as the density of probability that the two ends of the~DNA are in 
contact with one another. 
\item The $r$-dependent closure factor is 
\begin{equation}\label{defjr}
J(r, L) = \left[\int_0^{r} \!\!\! S(r', L) \; \mathrm{d}r'\right]/
\left(\frac{4}{3} \pi r^3\right). 
\end{equation} 
It gives the density probability for the two ends of the chain to stay within 
a sphere of radius~$r$. It is easy to check that~$J(r, L) \to J(L)$ 
when~$r \to 0$. For experimental convenience, units used for~$J(L)$ and~$J(r, L)$ 
are moles~per~liter: $1~\mathrm{nm}^{-3} \approx 1.66~\mathrm{mol}\cdot\mathrm{L}^{-1} %
\equiv 1.66~\mathrm{M}$. In these units $J(r, L)$~gives directly the concentration 
of one binding site in proximity of the other. 
\item Finally we consider the looping free energy cost 
\begin{equation}\label{defdeltag} 
\beta \Delta G(r, L) = - \ln{\! \left[J(r, L) \times \frac{4}{3} \pi r^3\right]}. 
\end{equation}
Note that this definition does not include the details of the geometry nor 
the affinities of the DNA/protein and protein/protein interactions. We actually 
assume all the sphere of radius~$r$ to be the reacting volume, \emph{i.e.} that 
the loop will form if the DNA~ends happen to be in this sphere.
\end{itemize}
%
%%%%%%%%%%%%%%%% 
%
Despite intense studies of the~WLC model no exact analytical expression is 
known for~$Q$ and the quantities of interest here, namely~$J$ and~$S$. However 
approximations expanding from the two limiting regimes 
(entropic~\cite{daniels, gobush} and elastic~\cite{wilhelm, shimada}) along with 
exact numerical computations are available. Hereafter, we have resorted to numerical 
as well as approximate analytical techniques~(SPA) for calculating the cyclization 
factor~$J$ and the probability of almost closed DNA~configurations. 
%
%%%%%%%%%%%%%%%%%%%%%%%%%%%%%%%%%%%%%%%%%%%%%%%%%%%%%%%%%%%%%%%%%%%%%%%%%%%%%% 
% 
\subsection{Numerical Calculation of the Probabilities~$P$,~$S$ and~$J$}\label{secnum} 
Our starting point for the calculation of the $z$~extension~PDF is the 
\textsc{Fourier}~representation of the \textsc{Dirac}~$\delta$-function 
in~(\ref{Q}) and~(\ref{P}), 
\begin{equation}\label{ft} 
P(z, L) = \int_{-\infty}^{+\infty} \frac{\mathrm{d}k}{2\pi} \; 
e^{+ i k z} \; Z \! \left(L, f = -i k/\beta\right). 
\end{equation} 
At fixed momentum~$k$ we are left with the calculation of the partition 
function~$Z(L, f)$ at (imaginary) force $f=-i k/\beta$. The path 
integral~(\ref{defz}) defining~$Z$ is interpreted as the evolution operator 
of a quantum system, the rigid rotator under an external imaginary field 
\begin{equation}\label{defz2}
Z(L, f) = \langle \mathrm{final} \vert 
\exp{\! \left[-L/A \times \widehat{H}(f)\right]} 
\vert \mathrm{initial} \rangle. 
\end{equation}
The entries of its hamiltonian~$\widehat H$ are easily expressed in the 
spherical harmonics~$\vert \ell, m \rangle$ basis: 
$\langle \ell, m \vert \widehat{H}(f) \vert \ell', m' \rangle = %
H_{\ell, \ell'}(f) \; \delta_{m, m'}$ with 
\begin{equation}\label{Hlm}
H_{\ell, \ell'}(f) = \frac{\ell(\ell+1)}{2} \, \delta_{\ell,\ell'} 
- \beta A f \, \frac{\ell \, \delta_{\ell', \ell-1} + 
\ell' \, \delta_{\ell, \ell'-1}}{\sqrt{(2\ell+1)(2\ell'+1)}}. 
\end{equation}
The entries of~$\widehat{H}$ do not depend on the azimuthal number~$m$ due 
to cylindrical symmetry around the force axis. Finally, integration over all 
initial and final orientations for the tangent vectors at the ends of the 
polymer chain selects~$\vert \mathrm{initial} \rangle = %
\vert \mathrm{final} \rangle = \vert 0, 0 \rangle$. \\ 
A recent paper~\cite{samuel} used~\textsf{Mathematica} to compute the vacuum 
amplitude~(\ref{defz2}) through a direct matrix exponentiation. We have instead 
used the~\textsf{Expokit} library~\cite{expokit} since it proves to be more 
accurate and faster for intensive numerical calculations. We truncated 
hamiltonian~(\ref{Hlm}) in a way the outcome is insensitive to the cut-off on 
the harmonics order. The~(inverse) \textsc{Fourier} transform~(\ref{ft}) is then 
handled by a Fast \textsc{Fourier} Transform~(FFT) algorithm~\cite{nr}. This 
task is in particular facilitated thanks to the inextensibility constraint which 
makes the distribution bandwidth limited. 
\begin{figure} [!h]
\includegraphics{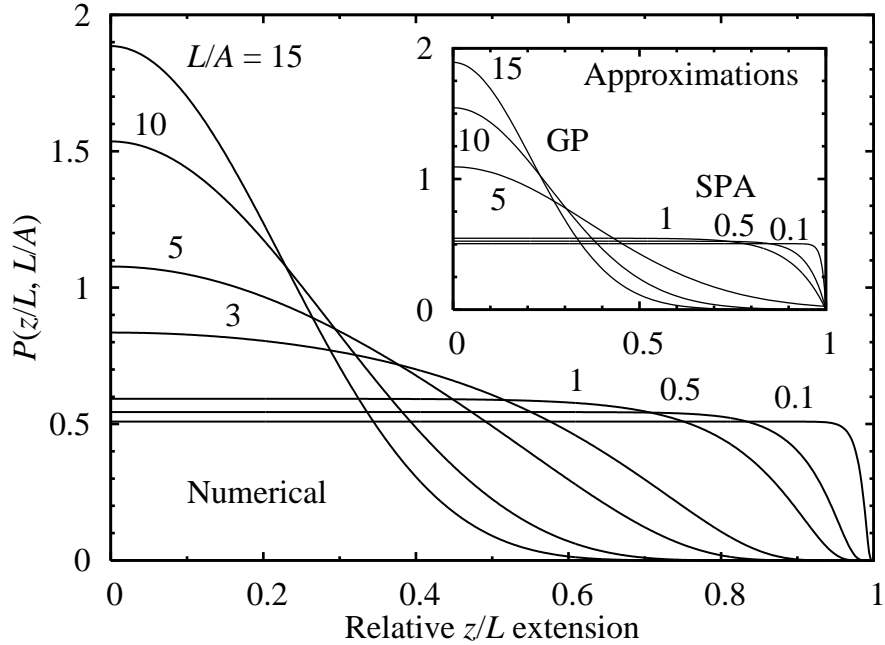} 
\caption{Numerical computation of the $z$~extension~PDF over a wide 
range of contour lengths~$L$. As expected the agreement with the~SPA prediction 
(see section~\ref{s:sp1}) improves as~$L$ decreases (upper bound~$L/A\lesssim 1$). 
The long~WLC behavior is caught by the GP approximation~\cite{yamakawa, kleinert} 
as soon as, say~$L/A\gtrsim 5$.} 
\label{f:Pz} 
\end{figure}
Our results for the $z$~extension~PDF are shown in Fig.~\ref{f:Pz}. The 
cross-over from the rigid elastic regime~$(L/A = 0.1, 0.5, 1)$ to the flexible 
entropic regime~$(L/A = 5, 10, 15)$ is clearly visible. Using~(\ref{relasp}) 
then gives us access to~$S$, the radial extension~PDF~(see Fig.~\ref{f:Sr}). 
The most probable value for the distance $r_\star$, switches from full 
extension~$r_\star \lesssim L$ for contour lengths~$L\lesssim A$ (elastic 
dominated regime) to the~GP most probable extension~$\sqrt{4LA/3}$ for longer 
ones~$L\gg A$ (entropic dominated regime). Also note that~$S$ always goes 
continuously to zero in the~$r\to L$ limit due to WLC~inextensibility. 
We have finally calculated the $r$-dependent closure factor~$J$ according 
to~(\ref{defjr}) by numerical integration of~$S$, and the looping free energy 
cost~$\Delta G$ defined in~(\ref{defdeltag}). \\
\begin{figure} [!h]
\includegraphics{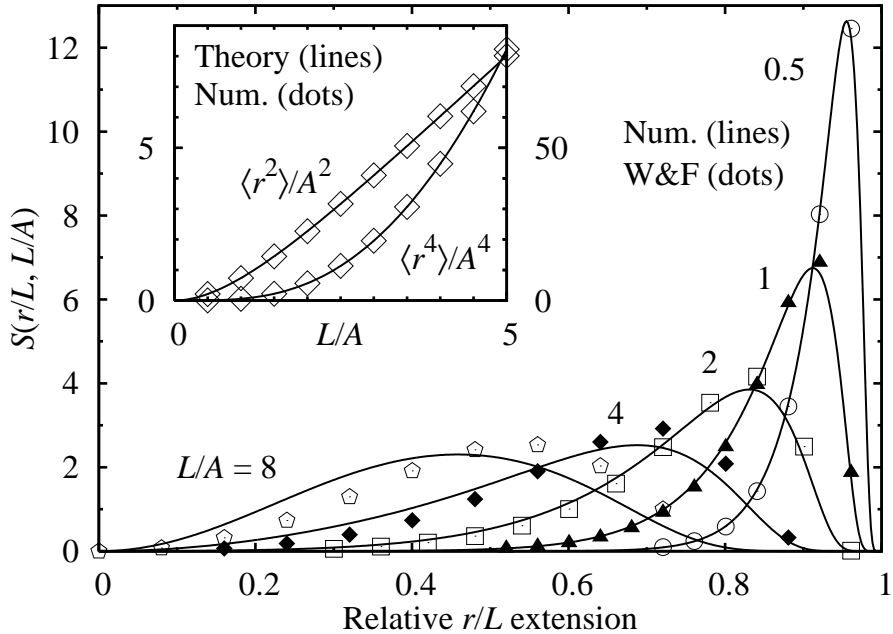}
\caption{Numerical computation of~$S$, the radial $r$~extension~PDF. 
The outcomes of the numerical calculations are tested against exactly 
known values for the first even 
moments~$\langle r^{2n}\rangle$ (inset)~\cite{yamakawa, kleinert}. The shape 
of~$S$ compares very well to the widely used~\textsc{Wilhelm} and~\textsc{Frey}~(W\&F) 
expansion~\cite{wilhelm}, valid up to~$L/A\lesssim 2$. Similar tests were 
achieved with other popular approximation schemes~\cite{focus} (data not shown).} 
\label{f:Sr} 
\end{figure}
Let us now discuss the numerical errors that could be important when calculating 
probabilities of rare events. Main sources of error are the hamiltonian~(\ref{Hlm}) 
truncation and the integration step to compute the~$r$ dependent closure 
factor~(\ref{defjr}). As mentionned above the cut-off on the harmonic order was 
systematically choosed in order convergence is observed. We have used $\ell = 50$ 
after having verified that the result is unchanged for $\ell = 100$. Concerning 
integration, limitation comes from the number of available data in the 
range~$0\le r'\le r$ which is directly related to the~$k$ sampling of the partition 
function~$Z$. For~$r=1$~nm the numerical integration is still reliable, but 
decreasing further~$r$ turns out to be critical. Other potential sources of error 
are negligeable. Indeed bandwidth limitation of~$P(z ; L)$ prevents any 
aliasing~\cite{nr} during~FFT~(\ref{ft}) and the derivative of $P$~(\ref{relasp}) 
can actually be skipped by an integration by part of (\ref{defjr}) to compute~$J$. 
Further hypothetical errors would then come from the~\textsf{Expokit} 
library~\cite{expokit} itself but its routines were coded to compute accurately 
matrix exponentials over a broad range of matrices~\cite{moler}. We have checked 
the numerical precision of our method by the comparison with the exact values for 
the first even moments  of $S(r)$ (Fig.~\ref{f:Sr} Inset); moreover as shown in 
Fig.\ref{f:Sr} $S(r)$ agrees with Wilhelm and Frey expansion for small~$L/A$ 
values; finally we will see in section~\ref{s:size} that the numerically 
calculated cyclization factor~$J(L)$ (\ref{defjr}) is in agreement with the 
\textsc{Shimada}-\textsc{Yamakawa} and Gaussian approximation results for 
respectively small and large~$L$ (Fig.~\ref{jyagaus}). 
%
%%%%%%%%%%%%%%%%%%%%%%%%%%%%%%%%%%%%%%%%%%%%%%%%%%%%%%%%%%%%%%%%%%%%%%%%%%%%%%%%%% 
%
\subsection{Saddle Point Approximation for~$J$}\label{s:sp1}
\begin{figure}
\includegraphics[height=12cm, angle=-90]{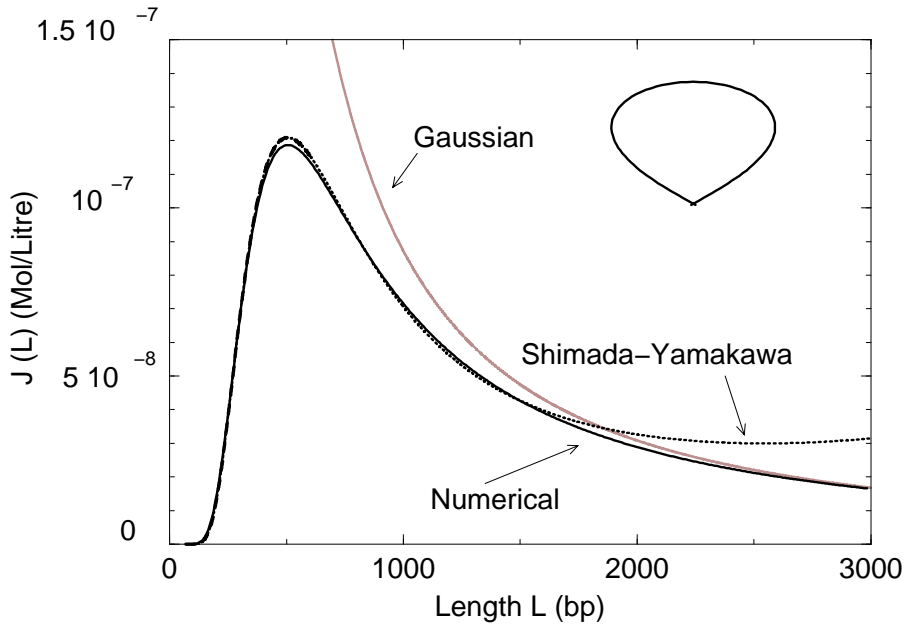}
\caption{Cyclization factor as a function of the DNA~length with: the gaussian 
model~(gray line); the WLC~model with the~\textsc{Shimada} and~\textsc{Yamakawa} 
formula~(dotted black line); the WLC~model with the numerical calculation~(full 
black line). The most probable length is 500~bp. Inset: the lowest bending energy 
configuration of a closed loop.} 
\label{jyagaus}
\end{figure}

\begin{figure}
\begin{tabular} {rl}
\includegraphics[height=8cm, angle=-90]{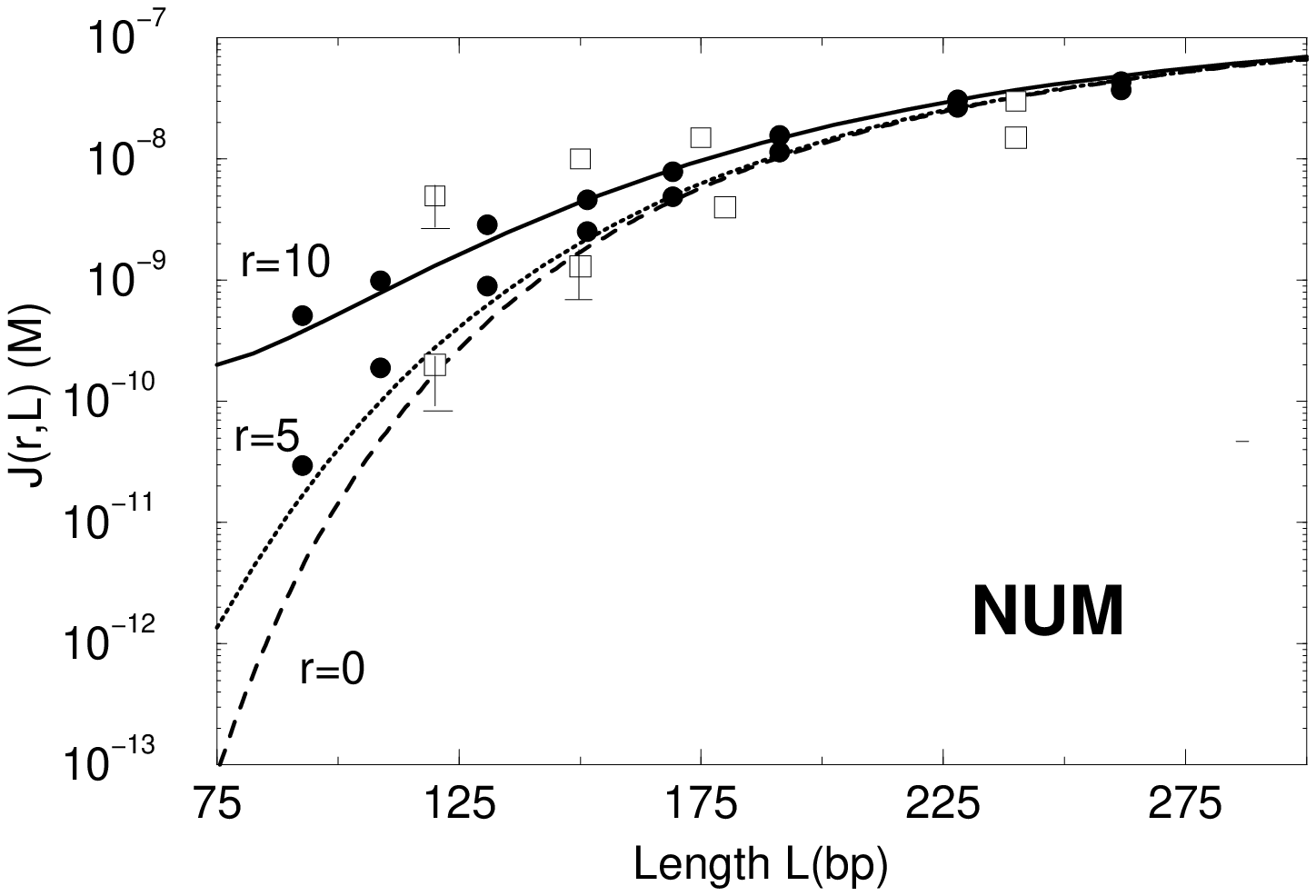} & 
\includegraphics[height=8cm, angle=-90]{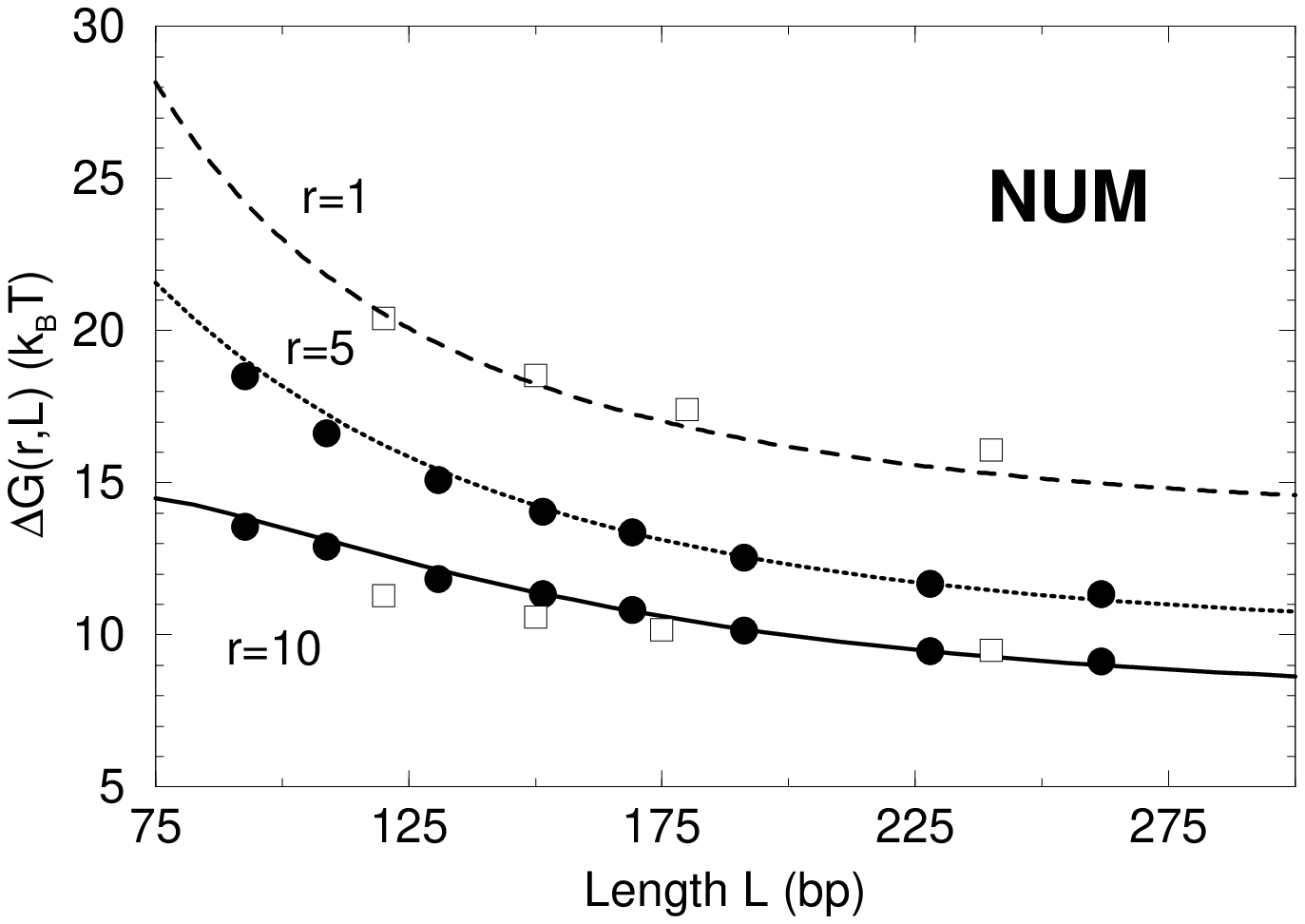} \\ 
\includegraphics[height=8cm, angle=-90]{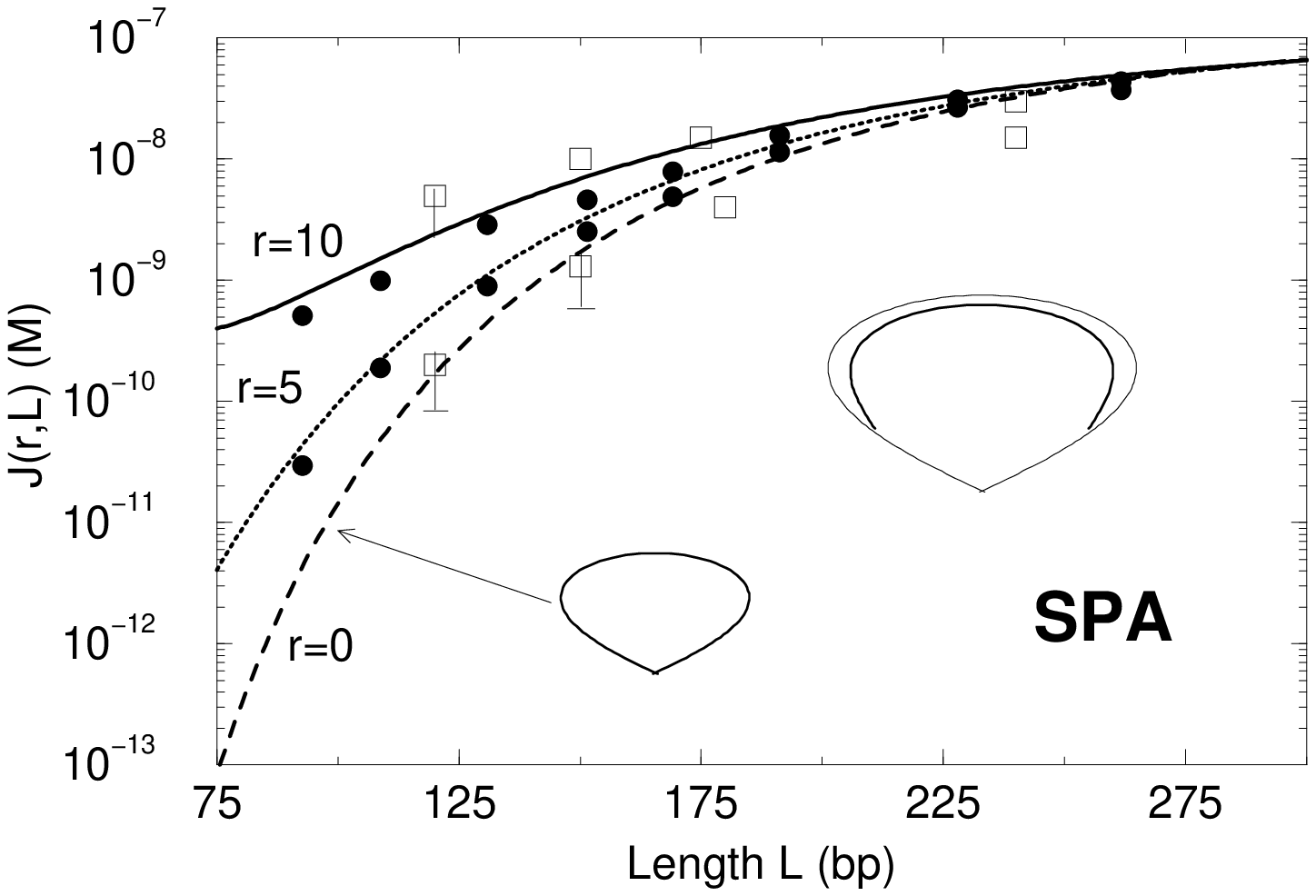} & 
\includegraphics[height=8cm, angle=-90]{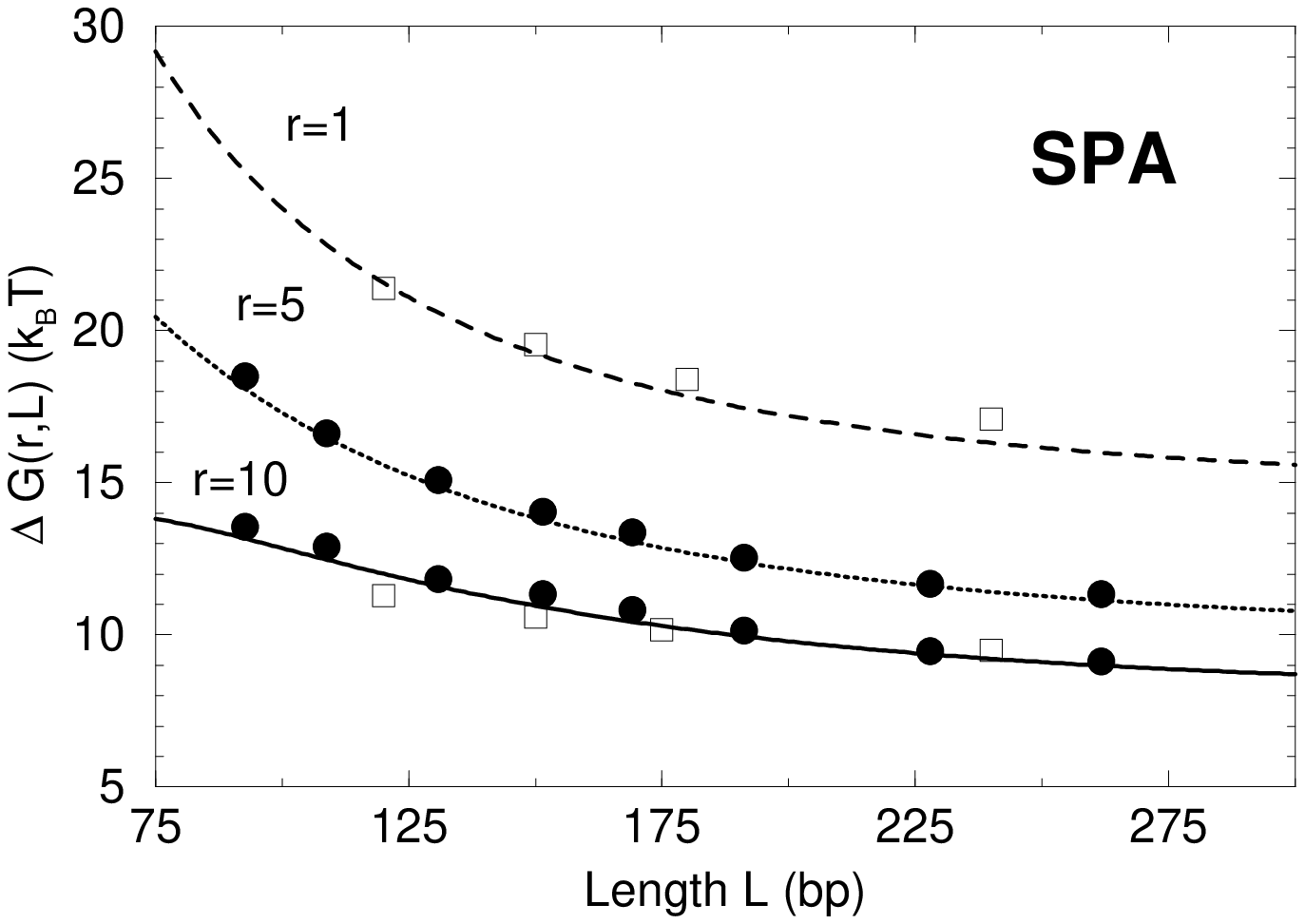} 
\end{tabular}
\caption{Closure factor~(left panel) and free energy~(right panel) with a 
protein bridge of sizes: $r = 1$~nm~(dashed line); $5$~nm~(dotted line); 
$10$~nm~(full line). The error bars are shown when they are larger than the 
symbol sizes. Top: numerical calculation of the constrained path integral. 
The~$r = 1$~nm curve coincides with the~$r = 0$ cyclization factor. Bottom: 
closure factor obtained by the extension of the~\textsc{Shimada} 
and~\textsc{Yamakawa} calculation, that includes~$r$. Theoretical results 
are in very good agreement with Monte Carlo simulations obtained 
by~\textsc{Podtelezhnikov} and~\textsc{Vologodskii} ~(filled circle, 
$\bullet$)~\cite{podtelezhnikov} and in fair agreement with brownian dynamics 
simulations obtained by~\textsc{Langowski} and~\emph{al.}~(empty 
square, $\Box$)~\cite{rippe95, merlitz}. Inset of the bottom left panel: 
lowest bending energy configurations for~100~bp and~$r=0$~(bottom) 
or~$r=10$~nm~(top), the closure of the~$r=10$~nm configuration is shown by 
a thin line.} 
\label{f:J}
\end{figure}

In addition to the exact numerical calculations detailed above, we have carried 
out approximate calculations based on a saddle point estimate of the partition 
function~(\ref{defz}). We follow the~\textsc{Shimada} and~\textsc{Yamakawa} 
calculation for the saddle point configuration~\cite{shimada}, extending it for 
an opened~DNA. The saddle point configuration for a closed loop is shown in 
Fig.~\ref{jyagaus} inset: it is a \emph{planar loop}. The tangent vector at 
position~$s$ along the chain is characterized by its angle~$\theta(s)$ with 
respect to the end-to-end extension~$\vec r$. The optimal configuration is 
symmetric with respect to the perpendicular~$\vec e_\perp$-axis while the 
half-length angle is~$\theta(L/2) = 180^\circ$. The initial angle~$\theta(0)$ 
is chosen by the minimization of the bending energy of the chain. The 
optimization gives rise to the following condition on the 
parameter~$x = \cos{\! \left[\theta(0)/4\right]}$ 
\begin{equation}
\left(1 + r/L\right) \; \widehat{K} \! \left(x^2\right) = 
2 \, \widehat{E} \! \left(x^2\right), 
\end{equation}
where~$\widehat{K} \! \left(x^2\right) = K\!\left(\frac{\pi}{2}, x^2\right)$ 
and~$\widehat{E} \! \left(x^2\right) = E\left(\frac{\pi}{2}, x^2\right)$ are 
the complete elliptical integrals of the first and second kinds 
respectively~\cite{abramowitz}. The corresponding elastic energy is 
\begin{equation}\label{f*} 
\beta \Delta E(r, L) = \frac {4}{L/A} \; \widehat{K}^2 \! \left(x^2\right) 
\times \left(2 x^2 - 1 + r/L\right). 
\end{equation}
The end-to-end extension~PDF is then approximated as 
\begin{equation}\label{kc}
Q(r, L) = C(r, L) \; \exp{\! \left[-\beta \Delta E(r, L)\right]}. 
\end{equation}
The prefactor~$C(r, L)$ should be calculated by taking into account quadratic 
fluctuations to the saddle point configuration. Since such calculation is quite 
involved we will actually only extend the~\textsc{Shimada} and~\textsc{Yamakawa} 
results which was computed considering fluctuations to a closed loop. 
In~$\mathrm{M} = \mathrm{mol}\cdot\mathrm{L}^{-1}$ units this reads 
\begin{equation}\label{c} 
C_\mathrm{SY}(L) = \frac{1.66}{A^3} \times \frac{112.04}{(L/A)^5} \, 
\exp{\! \left(0.246 \times L/A\right)}. 
\end{equation}
For this factor to fit the correct fluctuations~(to the opened loop) we have 
to consider the fluctuations to a fake closed loop of similar bending energy. 
Such a loop may be obtained by considering the optimal configuration of a loop 
of contour length~$(L + 2r)$ instead of~$L$, as shown in Fig.~\ref{f:J}~(bottom, 
inset). The choice of the factor~$2r$ derives from the following geometrical 
considerations: the closed saddle point configuration has an initial 
angle~$\theta(0) = 49.2^\circ$, the~$\Delta L$ closing the loop could be 
calculated for each value of~$r$ by requiring 
\begin{equation}
\int_0^{\Delta L} \!\!\!\!\!\! \cos{\theta(s)} \; \mathrm{d}s = r. 
\end{equation}
The angle~$\theta(s)$ increases slightly on the first part of the 
trajectory~$\theta(s) > \theta(0)$ and~$\Delta L > 1.53 \; r$. We have 
chosen~$\Delta L = 2r$ as an approximate value, this approximation has the 
advantage that it can be directly put in the fluctuation 
expression~$C(r, L) \approx C_\mathrm{SY}(L + 2r)$ in equation~(\ref{kc}) 
to obtain: 
\begin{equation}\label{qsp}
Q(r, L) = C_\mathrm{SY}(L + 2r) \; \exp{\! \left[-\beta \Delta E(r, L)\right]}
\end{equation}
where~$\Delta E$ is given in formula~(\ref{f*}) and~$C_\mathrm{SY}$ 
is given in formula~(\ref{c}). The validity of this approximation for the 
fluctuations prefactor was checked out by comparing~$J(r , L)$ obtained 
from~$Q(r, L)$ through formula~(\ref{defjr}), with the numerical results. 
The good agreement shown in Fig.~\ref{f:J} allows to obtain a semi-analytical 
formula for the loop probability with a finite interacting volume, which is 
valid for molecules of up to 2~kb~(kilo base pairs). 
% 
%%%%%%%%%%%%%%%%%%%%%%%%%%%%%%%%%%%%%%%%%%%%%%%%%%%%%%%%%%%%%%%%%%%%%%%%%%%%%
%
\section{Results: the effect of a Protein Bridge}\label{s:size} 
In Fig.~\ref{jyagaus} we show the cyclization factor~$J(L)$ for lengths~$L$ 
up to~3~kb. The most probable loop length is of about~$L_\star = 500$~bp that 
is~$L_\star/A\approx 3.5$~\cite{shimada, rippe95}. Both shorter and longer cyclized 
lengths are less probable: stiffness makes difficult the bending of smaller polymers 
while entropy makes longer polymers ends unlikely to meet. The numerical cyclization 
factor is compared with the saddle point calculation of \textsc{Shimada} 
and~\textsc{Yamakawa}~\cite{shimada} and with the gaussian polymer~(GP) model. The 
former is in good agreement with the numerical results for lengths smaller than 
about~1.5~kb while the latter works well for loops larger than~2.5~kb. \\
The effects of the finite size~$r$ of the protein bridge are displayed 
in~Fig.~\ref{f:J} which shows~$J(r, L)$~(on the left) and~$\Delta G(r, L)$~(on 
the right) for lengths~$L$ ranging from~75~bp~(\emph{i.e.}~25~nm) 
to~300~bp~(\emph{i.e.}~100~nm) and~$r$ respectively ranging from 1~nm to 10~nm. 
The numerical results (on the top ) are in good agreement with the SPA~results~(on 
the bottom). Note that for small lengths~$L$, $J(r, L)$~has a peak for 
the~$L \approx r$ event corresponding to rigid rod-like configurations. 
Fig.~\ref{f:J}  does not show this peak, occurring for~$r = 10$~nm at 
about~$L = 30$~bp~(or~$L=10$~nm) because we focus on the cyclisation events. 
As shown in this figure $r$~values ranging from~1~nm to~10~nm make no difference 
for the $r$-dependent closure factor, considering contour lengths~$L$ larger 
than~300~bp~(or~100~nm). The cyclization factor~$J(L)$ is evaluated as the 
limit~$r\to 0$ of~$J(r, L)$. In practice convergence is reached as soon as~$r$ 
is about one order of magnitude smaller than~$L$. In the range~$L > 75$~bp~(25~nm) 
$J(5~\mathrm{nm}, 150~\mathrm{bp})$ converges 
to~$J(1~\mathrm{nm}, 150~\mathrm{bp})$. On the other hand~$J(r, L)$ is 
considerably different from~$J(L)$ when~$r$ is of the same order of magnitude 
than~$L$. For example for loops of $L = 100$~bp~(34~nm) an end-to-end extension 
of~10~nm increases by two orders of magnitude the closure factor. Therefore 
proteins of size~$\approx 10$~nm are expected to produce drastic enhancements in 
looping short DNA~sequences. \\
In terms of energetics~(see Fig.~\ref{f:J}, right) cycling a~100~bp DNA~sequence 
costs~25~$k_\mathrm{B}T$ when the loop ends are required to stay within a sphere 
of radius~$r = 1$~nm. This cost decreases to~$13~k_\mathrm{B}T$ if the sphere has 
the typical protein size of~$r=10$~nm. For loop lengths larger than~300~bp the 
only difference in the three curves of Fig.~\ref{f:J}~(right), is a free energy 
shift due to the difference in the reacting volume. For instance, a reaction 
radius of~10~nm decreases the looping free energy 
of~$3 \times \ln(10) \approx 7~k_\mathrm{B}T$ with respect to a~1~nm reaction 
radius. \\
Our results for the closure factor and the looping free energy are compared in 
Fig.~\ref{f:J} with the Monte Carlo~(MC) and Brownian Dynamics~(BD) simulations 
results, obtained respectively by~\textsc{Podtelezhnikov} 
and~\textsc{Vologodskii}~\cite{podtelezhnikov}~(shown in the figure with filled 
circles, $\bullet$) and~\textsc{Langowski} 
and~\emph{al.}~\cite{rippe95, merlitz}~(displayed in the figure with empty 
squares, $\Box$). Numerical and SPA~data~(that for~$r=1$~nm converge to 
the~\textsc{Shimada} and~\textsc{Yamakawa} curve) are in better agreement with 
the~MC data than with the~BD ones; indeed numerical, SPA~and~MC data are obtained 
with a simpler model than BD~ones, which does not include twist rigidity and 
electrostatic effects. \\ 
Considerations about Lac~operon repression energetics will help us illustrating 
our results and compare them with other previous results. Expression of proteins 
enabling bacteria~\emph{E.~Coli} to perform the lactose metabolism can be prevented 
at the transcriptional level by cycling two different sequences~\cite{finzi}, 
the smallest one including the operon promoter. 
Let~$\mathrm{O}_1\mathrm{O}_3 = 76$~bp and~$\mathrm{O}_1\mathrm{O}_2 = 385$~bp 
denote these two resulting DNA~loops where the so-called 
operators~$\mathrm{O}_{1, 2, 3}$ actually are the small specific 
DNA~sequences~(10~bp or so) at which the tetrameric repressor protein~LacR can 
bind thus clamping the desired loop. Notice both processes are needed for efficient 
repression: despite~$\mathrm{O}_1\mathrm{O}_3$ contains the operon promoter its 
cyclization is much less probable to occur than the~$\mathrm{O}_1\mathrm{O}_2$ 
one~(see cyclization factor~$J(L)$ in Fig.~\ref{jyagaus}). The LacR~size is estimated 
from its cristallized stucture to~$r\approx 13$~nm~\cite{brenowitz}. In the \emph{in 
vitro} experiments many parameters are under control among which the operators sequence 
and location. The distance between the two operators, that defines the length of the 
DNA~loop has been fixed in~\cite{finzi, lia} to about~100~bp. Our results are in good 
agreement with the experimental measured stability of a~114~bp DNA~loop mediated by a 
LacR~protein, obtained by \textsc{Brenowitz} and~\emph{al.} in~1991~\cite{brenowitz}. 
By measuring the proportion of looped complexes present in a solution with respect to 
the unlooped molecules they obtained a looping free enegy 
of~$20.3 \pm 0.3~k_\mathrm{B}T$ to wich they associated a closure factor 
of~$8 \; 10^{-10}$~M. From Fig.~\ref{f:J}, the closure factor of a loop of~114~bp with 
a protein bridge of~$r = 10$~nm is~$J(10~\mathrm{nm}, 114~\mathrm{bp}) = 10^{-9}$~M to 
which we associate, from formula~(\ref{defdeltag}) a cyclization free 
enegy~$\Delta G(10~\mathrm{nm}, 114~\mathrm{bp}) = 12~k_\mathrm{B}T$. Note that the 
very good agreement between the closure factor contrasts with the bad agreement 
for the cyclization free energy. The latter could have been calculated considering 
a different reaction volume or it could also include the competition with 
configurations that do not allow the formation of a loop (see Fig.~2 
of~\cite{brenowitz}). To explain the high value found for the closure factor 
\textsc{Brenowitz} and~\emph{al.} already included the size of the protein in 
the analysis of their results by comparing their~$J$ result with the value 
expected for the cyclization probability of a free~DNA when the length of the 
protein is included in the size of the loop. \\
Another result on DNA~loop mediated by LacR protein has been obtained by 
\textsc{Balaeff} and~\emph{al.}~\cite{balaeff} who have numerically calculated 
the elastic energy of the $\mathrm{O}_1\mathrm{O}_3$~loop from a WLC~model 
also including: the twist rigidity, the short range electrostatic repulsion 
and the details of the LacR/DNA complex crystal structure. The elastic energy 
is estimated to~23~$k_\mathrm{B}T$ in~\cite{balaeff}, of which~$81 \%$ (that is 
$18~k_\mathrm{B}T$) due to the bending and~$19\%$ due to the unwinding. The 
bending energy of~$18~k_\mathrm{B}T$ is to be compared to the saddle point 
energy~(\ref{f*}) of $15~k_\mathrm{B}T$ for a~75~bp loop with an end-to-end 
extension of~$r = 10$~nm. The corresponding free energy of loop formation 
obtained by~(\ref{qsp}) and after integration over the reacting 
volume~(\ref{defdeltag}) 
is~$\Delta G(10~\mathrm{nm}, 75~\mathrm{bp}) = 14~k_\mathrm{B}T$ (see 
Fig.~\ref{f:J}, bottom/right). For a~400~bp loop since LacR is 
only~$\approx 10 \%$ of the loop length, its size is expected to play a less 
important role. Indeed the cyclization probability does not depend 
on~$r < 10$~nm and the free energy of forming such a loop decreases 
from~$15~k_\mathrm{B}T$ for~$r=1$~nm to~$8~k_\mathrm{B}T$ for~$r=10$~nm only 
because the reaction volume increases by a 
factor~$3 \times \ln{(10)} \approx 7~k_\mathrm{B}T$. 
%
%%%%%%%%%%%%%%%%%%%%%%%%%%%%%%%%%%%%%%%%%%%%%%%%%%%%%%%%%%%%%%%%%%%%%%%%%%%
%
\section{Effect of the Presence of a Kink}\label{s:kink} 
The previous protein mediated DNA~looping modelization (section~\ref{s:size}) 
assumes that the only intervening proteins are the ones clamping the loop ends 
(\emph{e.g.}, the Lac~operon repressor~LacR). Actually regulation phenomena 
involve several proteins which may bind along the whole~DNA. Indeed naked~DNA 
situations barely exist \emph{in vivo}. For instance single molecule 
manipulations~\cite{lia} have shown that efficient Gal~operon repression needs 
a stiffness loss of the 113~bp DNA portion to be looped. The HU~protein produces 
such loss by kinking the sequence. 
%
%%%%%%%%%%%%%%%%%%%%%%%%%%%%%%%%%%%%%%%%%%%%%%%%%%%%%%%%%%%%%%%%%%%%%%%%%%%% 
% 
\subsection{Numerical Calculation of~$J$ for a Bridged and Kinked Loop}
Such stiffness loss may be taken into account in an effective way by kinking 
the~WLC at half-length~$L/2$. Let us call $\theta _-$ and $\theta_+$ the angles
of the~DNA just before and after the kink respectively. We assume that the kink 
plane is vertical and choose it to define the origin of the azimuthal 
angle~$\phi$: $\phi_- = \phi_+ = 0$. Using the quantum language of 
section~\ref{secnum}, we replace the calculation of the evolution 
operator~$Z(L, f)$ (\ref{defz2}) with its kinked counterpart
\begin{eqnarray}
Z_\mathrm{kinked}(L, f) & = & 
\langle \mathrm{final} \vert 
\exp{\! \left[-\frac{L}{2A} \, \widehat{H}(f)\right]} 
\vert \theta _+, \phi_+ \rangle 
\times
\langle \theta_-, \phi_- \vert 
\exp{\! \left[-\frac{L}{2A} \, \widehat{H}(f)\right]} 
\vert \mathrm{initial} \rangle 
\nonumber \\ 
 & = & 
\sum_{\ell, \ell'} 
\langle 0, 0 \vert 
\exp{\! \left[-\frac{L}{2A} \, \widehat{H}(f)\right]} 
\vert \ell, 0 \rangle 
\, 
\langle \ell', 0 \vert 
\exp{\! \left[-\frac{L}{2A} \, \widehat{H}(f)\right]} 
\vert 0, 0 \rangle 
\times Y_\ell^0 (\theta_+, 0) \, Y_{\ell'}^0 (\theta_-, 0) 
\end{eqnarray}
where we have used the change of basis from angles to spherical harmonics 
\begin{equation}
\vert \theta_\pm, \phi_\pm = 0 \rangle = 
\sum_{\ell \ge 0} Y_\ell^0(\theta_\pm, 0) \vert \ell, 0 \rangle. 
\end{equation}
Although calculations are a little bit more involved the evaluation scheme 
for the cyclization factor~$J$ remains unchanged in its 
principle~(section~\ref{secnum}). We now have ``$\vert \ell\ne 0,0 \rangle$ 
elements'' corresponding to particular orientations arriving at~($-$) and 
leaving from~($+$) $s = L/2$. The kink angle~$\kappa$ is intuitively defined 
from these WLC~tangent vectors at half-length (in spherical coordinates) 
through 
\begin{equation}
\kappa = \theta_+ - \theta_-. 
\end{equation}
\begin{figure}
\begin{tabular} {@{\hspace{-2.5cm}}r@{\hspace{-8.4cm}}c%
@{\hspace{-8.4cm}}c@{\hspace{-8.4cm}}l} 
\includegraphics[height=9cm]{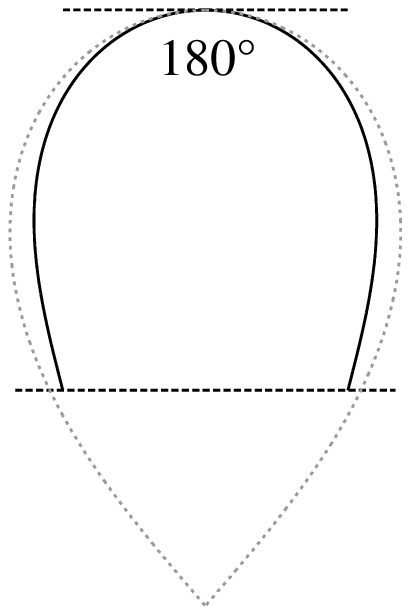} & 
\includegraphics[height=9cm]{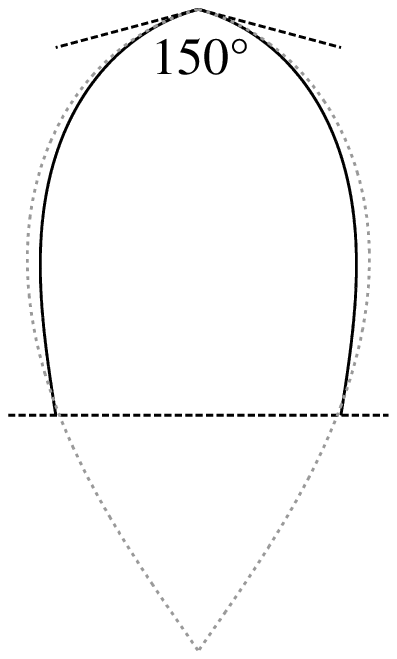} & 
\includegraphics[height=9cm]{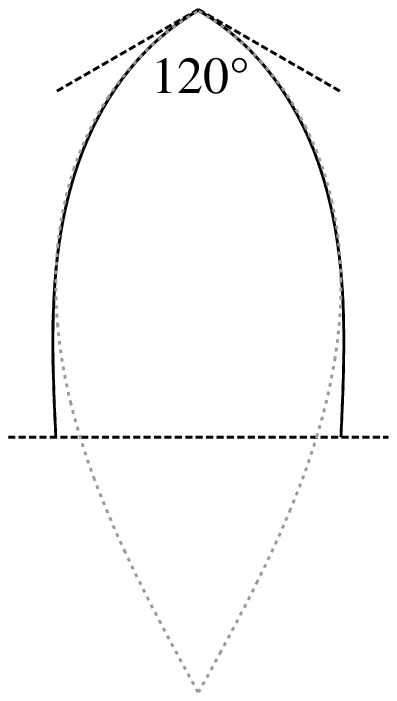} & 
\includegraphics[height=9cm]{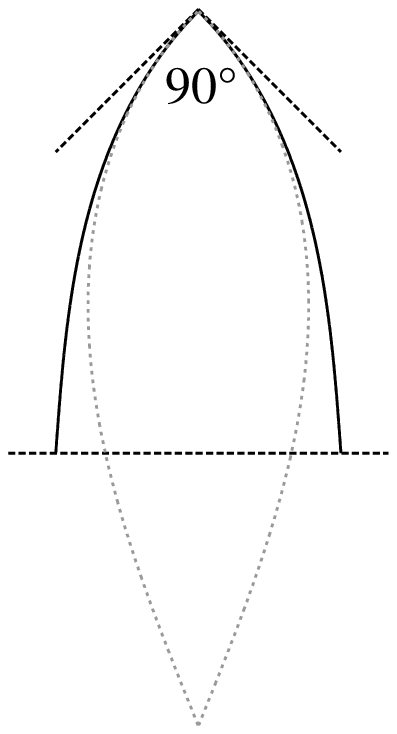} 
\end{tabular}
\caption{Lowest bending energy configurations for a~100~bp DNA~loop with an 
end-to-end extension~$r = 10$~nm and kinks~$\kappa = 90^\circ$, $120^\circ$, 
$150^\circ$, $180^\circ$ respectively. The gray configurations are the closed 
loops  used in the calculation of the 
fluctuation.} 
\label{confkink}
\end{figure}
\begin{figure}
\begin{tabular} {rl} 
\includegraphics[height=8cm, angle=-90]{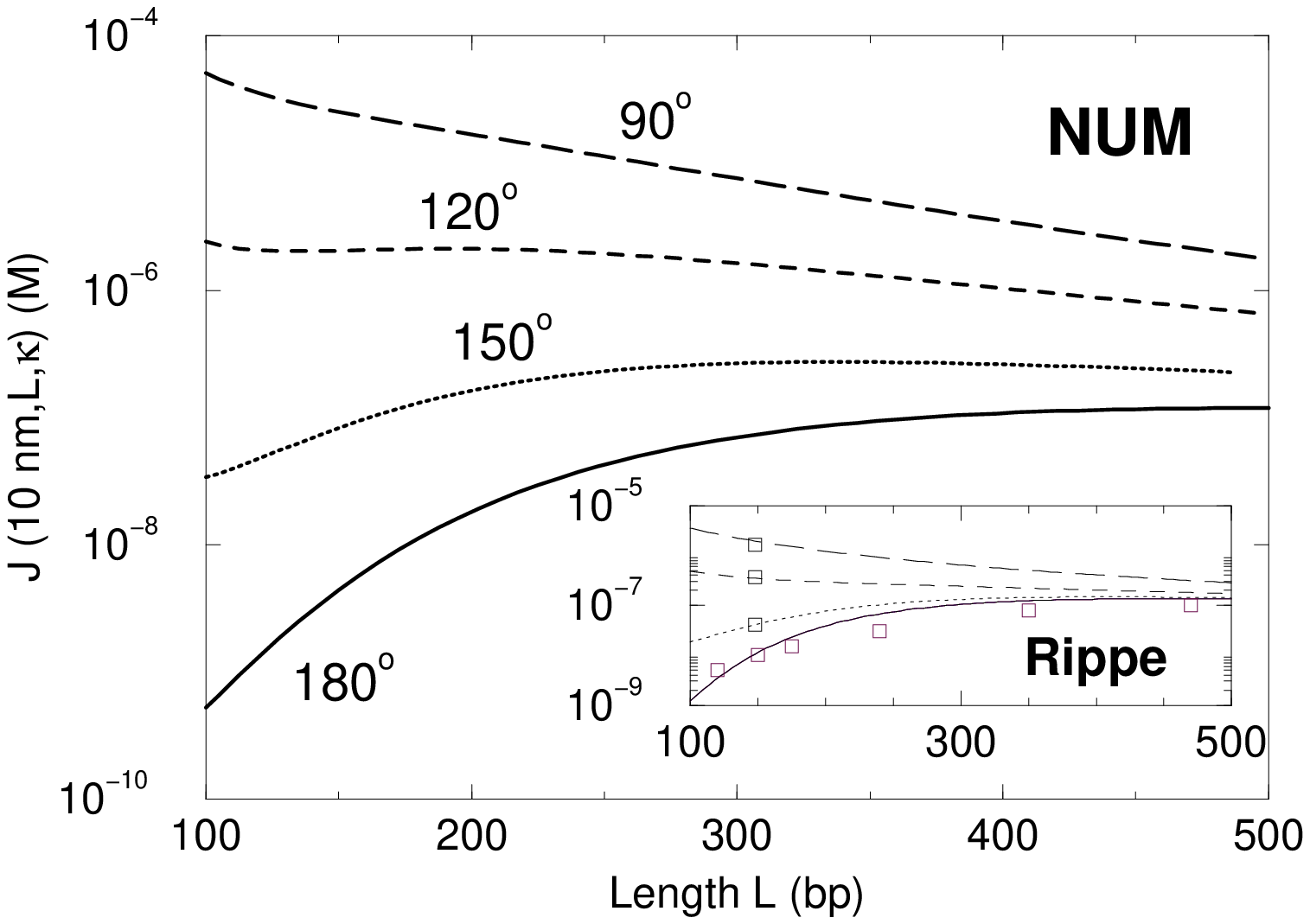} &
\includegraphics[height=8cm, angle=-90]{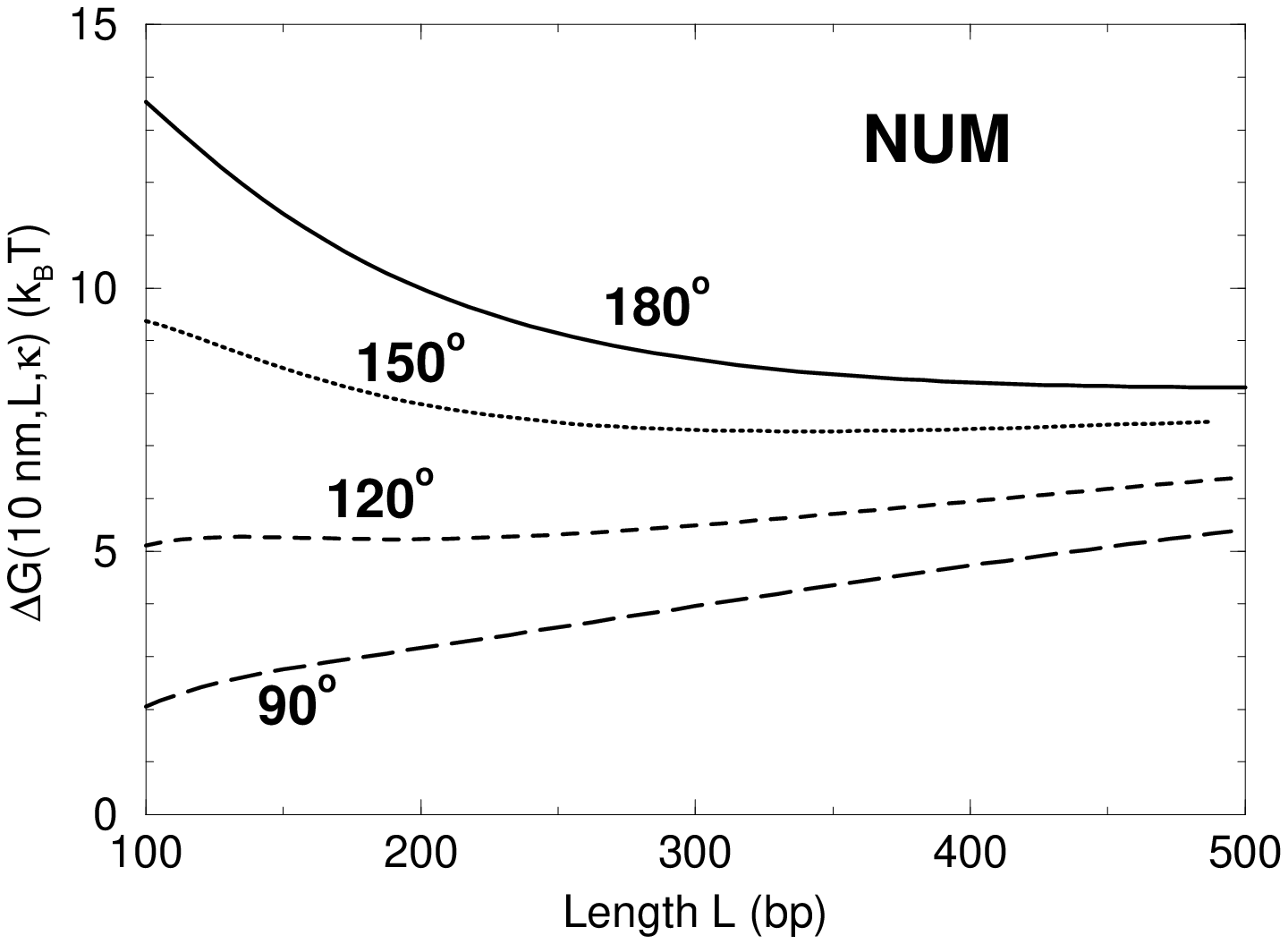} \\
\includegraphics[height=8cm, angle=-90]{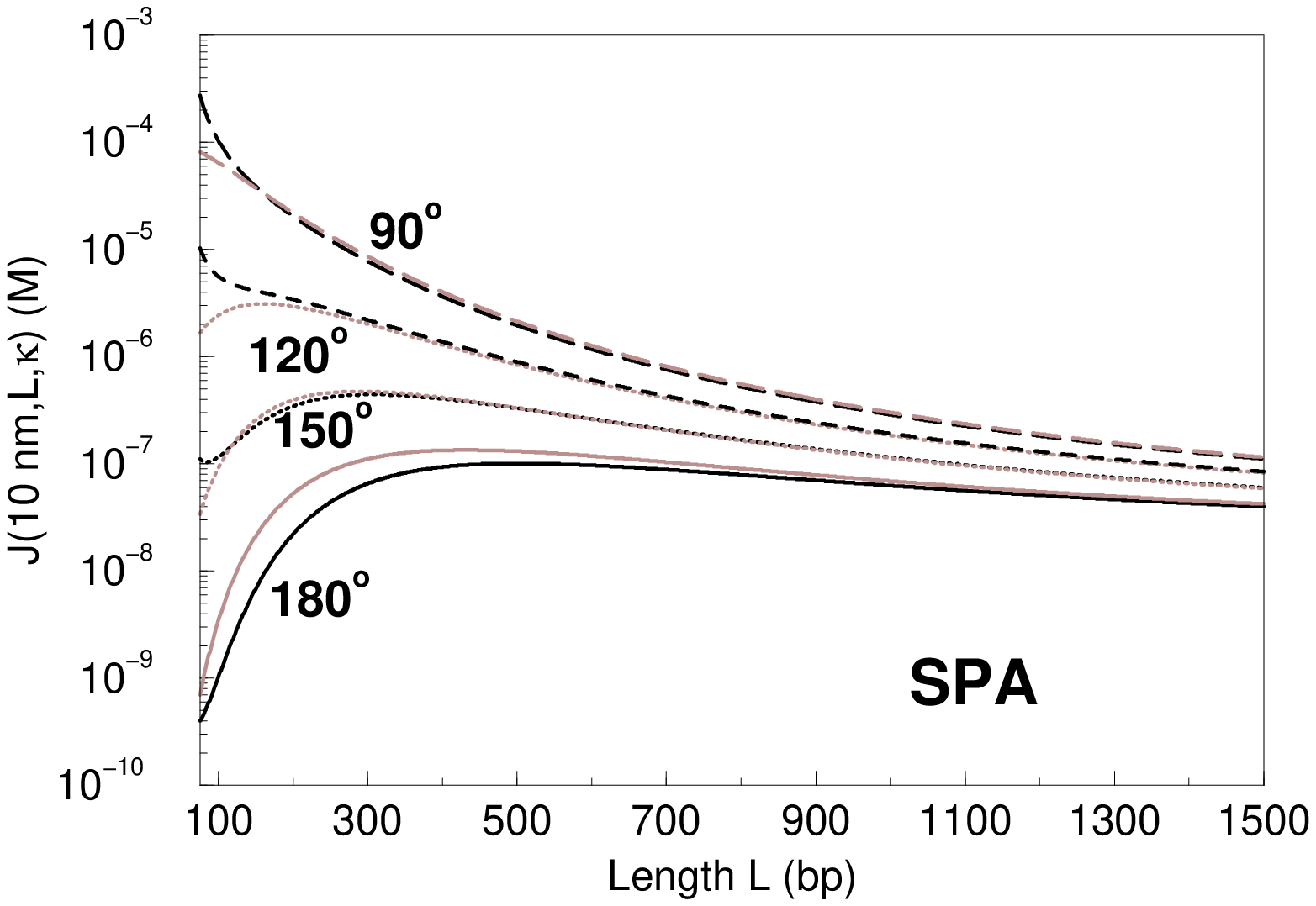} &
\includegraphics[height=8cm, angle=-90]{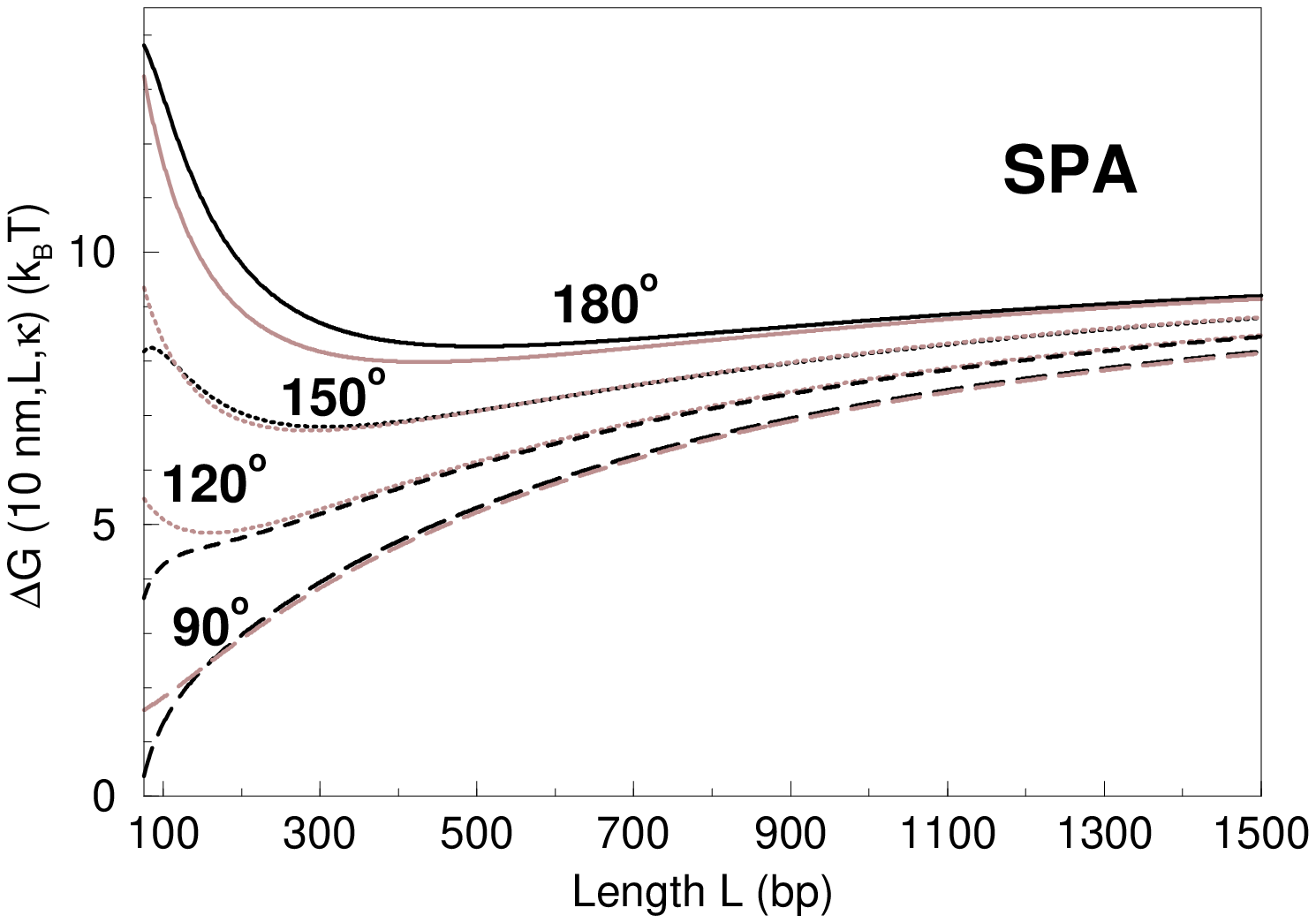}
\end{tabular}
\caption{Closure factor (left) and free energy for a loop with~$r = 10$~nm 
and a kink~$\kappa$ in the middle of the chain: $\kappa=180^\circ$~(full lines); 
$\kappa = 150^\circ$~(dotted lines); $\kappa = 120^\circ$~(dashed line); 
$\kappa = 90^\circ$~(long dashed lines). Top: numerical calculation of the 
constrained path integral. Bottom: extension of the~\textsc{Shimada} 
and~\textsc{Yamakawa} calculation~(black lines) and approximate formula~(gray 
line) given in the text~(\ref{for}). Numerical results are in very good 
agreement with the SPA~approximation and the approximate formula~(\ref{for}).
Inset: brownian dynamics simulations point obtained by the~\textsc{Langowski} 
and collaborators~(empty square, $\Box$)~\cite{rippe95, merlitz} , fitted by a 
simple formula by~\textsc{Rippe}~\cite{rippe00}.} 
\label{jr10}
\end{figure}
% 
%%%%%%%%%%%%%%%%%%%%%%%%%%%%%%%%%%%%%%%%%%%%%%%%%%%%%%%%%%%%%%%%%%%%%%
% 
\subsection{Saddle Point Approximation of~$J$ for a Bridged and Kinked Loop} 
The saddle-point calculation of section~\ref{s:sp1} can be straightforwardly 
extended to the case of a kinked loop. In Fig.~\ref{confkink} we show the 
configurations with the lowest bending energy for a~100~bp DNA~loop with an 
end-to-end extension~$r=10$~nm. The kink is accounted for by a bending angle 
in the middle of the chain~$\kappa = 2 \times \theta(L/2) - \pi$ \emph{a priori} 
different from the previous trivial value~$\kappa=\pi$, ranging from~$150^\circ$ 
to~$90^\circ$. We introduce the phase~$\psi = \arcsin{\! \left[\sin{\! %
\left(\frac{\pi-\kappa}{4}\right)}/x\right]}$. The parameter~$x$ is now obtained 
from equation 
\begin{equation}
\left(1 + r/L\right) \; 
\left[\widehat{K} \! \left(x^2\right) - K \! \left(\psi, x^2\right)\right] = 
2 \left[\widehat{E} \! \left(x^2\right) - E \! \left(\psi, x^2\right)\right] 
\end{equation}
and the total bending energy is 
\begin{equation}\label{delk}
\beta \Delta E(r, L, \kappa) = \frac{4}{L/A} \left[\widehat{K} \! \left(x^2\right) 
- K \! \left(\psi, x^2\right)\right]^2 \left(2 x^2 - 1 + r/L\right). 
\end{equation}
In analogy with~(\ref{kc}) we obtain the end-to-end extension~$\vec r$~PDF 
\begin{equation}\label{jrlk}
Q(r, L, \kappa) = C_\mathrm{SY}(L + 2r) \, 
\exp{\! \left[-\beta \Delta E(r, L, \kappa)\right]}, 
\end{equation}
where~$C_\mathrm{SY}$ is given in~(\ref{c}), from wich we 
calculate~$J(r, L, \kappa)$ through formula~(\ref{defjr}). Note that~(\ref{jrlk}) 
reduces to the loop probability given in~\cite{sankararaman} for a closed and 
kinked~DNA. Again the good agreement obtained with numerical results allows us 
to establish a semi-analytical formula for the loop probability with a finite 
interacting volume and kink. 
%
%%%%%%%%%%%%%%%%%%%%%%%%%%%%%%%%%%%%%%%%%%%%%%%%%%%%%%%%%%%%%%%%%%%%%%%%% 
%
\subsection{Results for a Kinked and Bridged Loop}\label{ss:kink}
\begin{figure}
\includegraphics[height=12cm, angle=-90]{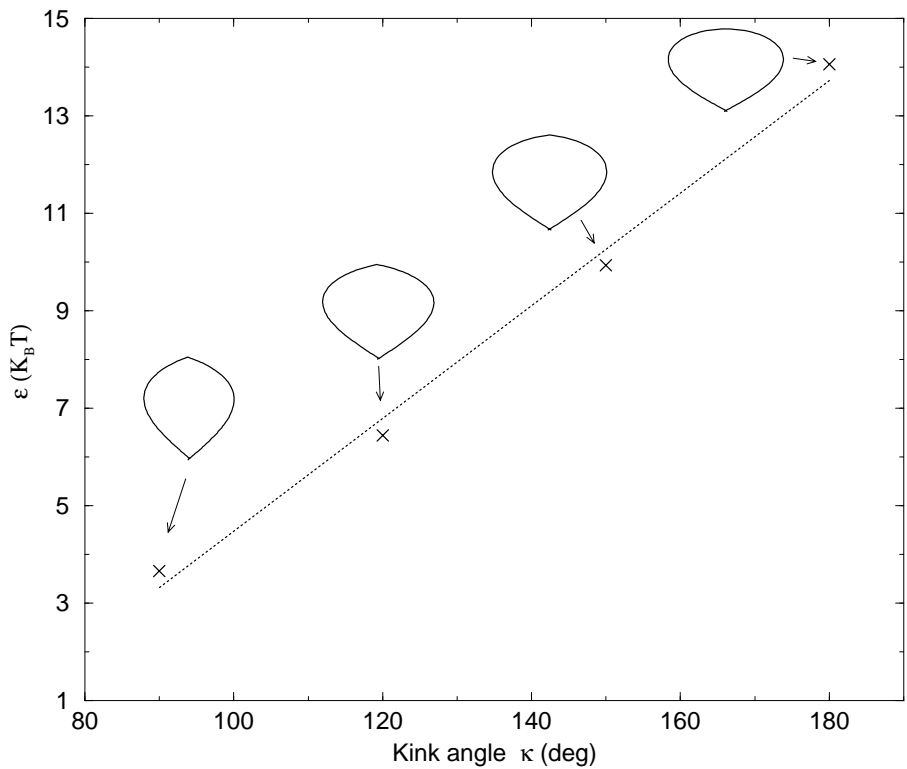}
\caption{The ``$\times$'' points: saddle point energy in units of~$L/A$~(that 
is~$\varepsilon = \Delta E \times L/A$) for the saddle point 
configurations~(also displayed in the figure) with kink angles 
of~$\kappa = 90^\circ$, $120^\circ$, $150^\circ$, $180^\circ$. Dotted line: 
linear interpolation used in formula~(\ref{for}).}
\label{eangoli}
\end{figure}

In Fig.~\ref{jr10} results for the looping probability density~(left) and free 
energy (right) are shown for a typical end-to-end extension~$r=10$~nm and 
kinks~$\kappa = 150^\circ$, $120^\circ$, $90^\circ$. The curve with no 
kink~($\kappa = 180^\circ$) is also shown for comparison. The numerical results 
on the top of the figure are in very good agreement with
 the extension of the~\textsc{Shimada} 
and~\textsc{Yamakawa} saddle point approach on the bottom of the figure~(black 
lines). In Fig.~\ref{jr10}~(inset) we show the results obtained by BD~simulations 
by~\textsc{Langowski} and~\emph{al.}~\cite{rippe95, merlitz} 
for~$J(10~\mathrm{nm}, L, \kappa)$~(empty squares, $\Box$) fitted 
by~\textsc{Rippe}~\cite{rippe00} with a simple formula containing one fitting 
parameter for each curve. The curves in the inset of Fig.~\ref{jr10} reproduce the 
same  behavior with~$L$ and~$\kappa$ of our numerical~(top of Fig.~\ref{jr10}) or 
SPA~curves~(bottom of Fig.~\ref{jr10}). The numerical gap with BD~results increases 
for large kinks: at~$\kappa =90^\circ$ our closure factor is ten times larger than 
the closure factor obtained by BD~simulations. The electrostatic and twist rigidity effect 
could indeed play a more important role when the chain is kinked. Note that the 
lengths range of the numerical results are from~100~bp to~500~bp, 
while the lengths range of the saddle point results is from~75~bp to 1500~bp. 
The lengths range of the~SPA is limited by the validity of the approximation 
$\approx 1500$~bp shown in Fig.~\ref{jyagaus}. As an example the closure 
factor~$J(10~\mathrm{nm}, 113~\mathrm{bp}, \kappa)$ of a~113~bp fragment, 
obtained by the numerical calculations, increases from the value of~$10^{-9}$~M 
for a non kinked loop~($\kappa=180^\circ$) to~$4 \, 10^{-8}$~M, $2 \, 10^{-6}$~M 
and~$4 \, 10^{-5}$~M for respectively~$\kappa = 150^\circ$, $120^\circ$ 
and~$90^\circ$. The corresponding looping free energy is 
$\Delta G(10~\mathrm{nm}, 113~\mathrm{bp}, \kappa) = 13~k_\mathrm{B}T$, 
$9~k_\mathrm{B}T$, $5~k_\mathrm{B}T$ and $2.5~k_\mathrm{B}T$ 
for~$\kappa = 180^\circ$, $150^\circ$, $120^\circ$ and $90^\circ$ respectively. 
With the saddle point approach we find similar results: 
$J(10~\mathrm{nm}, 113~\mathrm{bp}, \kappa) = 2 \, 10^{-9}$~M, $10^{-7}$~M, 
$5 \, 10^{-6}$~M and $8 \, 10^{-5}$~M while 
$\Delta G(10~\mathrm{nm}, 113~\mathrm{bp}, \kappa) = 11~k_\mathrm{B}T$, 
$8~k_\mathrm{B}T$, $4.4~k_\mathrm{B}T$ and $1.7~k_\mathrm{B}T$. As it is shown 
in Fig.~\ref{jr10} the presence of the kink become irrelevant for DNA~segment 
larger than about~1500~bp. The example of a 113~bp DNA~segment has been chosen 
to compare the results with the single molecule experiments on the~GalR mediated 
loop of an~$\approx 113$~bp DNA~portion between the two operators. From the 
kinetics of the loop formation mediated by the~GalR and~HU proteins, \textsc{Lia} 
and~\emph{al.}~\cite{lia} have deduced a looping free energy of~$12~k_\mathrm{B}T$, 
that with respect to our values should correspond to a kink angle of more 
than~$150^\circ$ or to an end-to-end extension smaller than~$r= 10$~nm. Another 
significant change in the  cyclization probability is that the stiffness loss 
induced by~$\kappa$ reduces the most probable loop length from~500~bp for a non 
kinked~DNA to 340~bp and 190~bp~(from the numerical calculation) or to 300~bp 
and 150~bp (from SPA) for kinks of respectively $150^\circ$ and $120^\circ$. 
Note that for a kinked loop with an end-to-end extension~$r$ the minimal 
length~$L_0$ corresponding to the rigid rod-like configuration fulfills the 
relation~$L_0 \sin(\kappa/2) = r$ and therefore it is of~$\approx 42$~bp 
for~$\kappa = 90^\circ$ instead of~$\approx 30$~bp for~$\kappa = 180^\circ$. 
for~$\kappa = 90^\circ$ the most probable loop length is the rigid \emph{kinked} 
rod-like configuration of the two half-DNA portions. To catch both kink and protein 
bridge effects in a simple formula, we have calculated the cyclization factor with 
the extension of~\textsc{Shimada} and~\textsc{Yamakawa} formula for a kinked closed 
loop of length~$(L + 2r)$. This approach is similar to what was suggested en~1991 by 
\textsc{Brenowitz} and~\emph{al.}~\cite{brenowitz} to interpret their experimental 
data, \emph{i.e.} to directly consider the protein as part of the length of the loop. 
A linear fit~\cite{nr} of the bending energy~(\ref{delk}) for the optimal 
closed configuration~($r = 0$) in the presence of a kink (expressed in 
degrees):~$\beta \Delta E(r = 0, L, \kappa) \approx \left(-7.1 %
+ 0.1155 \, \kappa\right)/(L/A)$, is shown in Fig.~\ref{eangoli}. It gives 
the following approximated formula for the closure factor as a function of the 
protein size~$r$, the length~$L$ of the~DNA, and the kink angle~$\kappa$ 
\begin{equation}\label{for}
J_\mathrm{approx} (L, r, \kappa) = C_\mathrm{SY}(L + 2r) \; 
\exp{\! \left[\frac{7.1 - 0.1155 \, \kappa}{(L+2r)/A}\right]}
\end{equation}
where~$C_\mathrm{SY}$ is given in~(\ref{c}). Notice the integration 
step~(\ref{defjr}) has been skipped since it does not make any significant 
difference. Formula~(\ref{for}) allows us to obtain a simple prediction for 
the loop probability in presence of a kink in the middle 
of the sequence and a finite separation between the extremities. As shown in 
Fig.~\ref{jr10} this formula~(gray lines) is in good agreement 
with the loop probability obtained with the exact calculation of the saddle point 
energy of the open configuration~(full line). In particular Fig.~\ref{jr10} 
shows that for kink angles in the range~$90^\circ < \kappa < 150^\circ$ this 
simple formula works remarkably well for lengths~$L$ larger than about $5r$, 
that is 150~bp for $r=10$~nm. For smaller lengths the optimal configuration 
is more a rigid rod-like and cannot be approximated by a closed loop. Similar 
simple formulas that includes a kink angle~$\kappa$ and a finite end-to-end 
distance~$r$ in an effective way have also been written down by~\textsc{Rippe} 
or~\textsc{Ringrose} to fit their brownian dynamics simulation~\cite{rippe00} 
or experimental data~\cite{ringrose}, but these formulas contain a parameter 
that must be fitted for each values of~$r$ and~$\kappa$ from the data points
(Fig.~\ref{jr10}). 
%
%%%%%%%%%%%%%%%%%%%%%%%%%%%%%%%%%%%%%%%%%%%%%%%%%%%%%%%%%%%%%%%%%%%%%%%%%%%%%%%%%%% 
%
\section{Conclusion}\label{s:conclusion} 
We performed both numerical and analytical calculations of the closure factor~$J$, 
even in the presence of a protein bridge and of a protein-mediated kink. More 
precisely we have numerically calculated the path integral of the WLC~polymer 
model under the constraints of a fixed end-to-end distance~$r$ and a kink~$\kappa$ 
in the middle of the DNA~portion. Analytically we have extended 
the~\textsc{Shimada} and~\textsc{Yamakawa} saddle point approximation~\cite{shimada} 
to the case of a bridged and kinked loop. We have seen that the formation of DNA 
loops is significantly sensitive to the size of the protein bridge when this 
size~$r$ is more than~$10\%$ of the loop length~$L$, that is~300~bp~(or~100~nm) 
for a typical protein bridge size of~$r=10$~nm. To give an example, the closure 
factor for a 100~bp DNA~segment increases 
from~$J(100~\mathrm{bp}, 0) \approx 10^{-11}$~M 
to~$J(100~\mathrm{bp}, 10~\mathrm{nm}) \approx 10^{-9}$~M. Correspondingly, 
looping free energy decreases from~$\Delta G(100~\mathrm{bp}, 0) = 24~k_\mathrm{B}T$ 
to~$\Delta G(100~\mathrm{bp}, 10~\mathrm{nm}) = 13~k_\mathrm{B}T$. A kink ranging 
from~$150^\circ$ to~$90^\circ$ produces a significant increase of~$J$ for 
DNA~fragments of lengths up to about~2500~bp. For instance the closure factor for 
a 100~bp DNA~segment with a protein bridge~$r = 10$~nm and a kink of~$90^\circ$ 
is~$J(100~\mathrm{bp}, 10~\mathrm{nm}, 90^\circ) \approx 10^{-4}$~M, and the 
corresponding looping free energy~$\Delta G(100~\mathrm{bp}, 10~\mathrm{nm}, %
90^\circ) \approx 2~k_\mathrm{B}T$. A kink also changes the most probable loop 
length from 500~bp~(no kink) to about 175~bp for a kink of~$120^\circ$, 
going to the rigid \emph{kinked} rod-like configuration for smaller~$\kappa$ values. 
This is an interesting mechanism because the loop lengths implied in \emph{in vivo} 
DNA~processing by proteins spread on a large lengths range. Our results were 
compared to previous analytical approximations~(in particular the results of the 
gaussian model, the~\textsc{Wilhelm} and~\textsc{Frey} expansion~\cite{wilhelm} 
and the~\textsc{Shimada} and~\textsc{Yamakawa} formula~\cite{shimada}) and 
numerical calculations~(in particular the Monte~Carlo simulations data obtained 
by~\textsc{Podtelezhnikov} and~\textsc{Vologodskii}~\cite{podtelezhnikov}, 
and the brownian dynamics simulations data obtained by \textsc{Langowski} 
and~\emph{al.}~\cite{rippe95, merlitz, rippe00})  as well as 
experimental results~\cite{brenowitz, finzi, lia} (\emph{e.g.}, the ones obtained 
by \textsc{Lia} and~\emph{al.} on the looping dynamics mediated by the~Gal and~HU 
proteins). Finally a simple formula~(\ref{for}) including both the protein bribge 
and kink effects has been proposed. This formula has the advantage of not containing 
adjustable parameters with respect to the existing formulas that include both these 
effects~\cite{rippe00}. \\ 
Still many effects omitted in this work can be included without significant changes 
in the numerical algorithm. The kink we considered is actually permanent~(that is 
not thermally excited), site specific~(at half-length) and rigid~($\kappa$~fixed). 
Although this rigidity seems relevant to most protein bindings to~DNA at first 
glance~\cite{popov}, it was pointed out in~\textsc{Yan} and~\emph{al.} 
works~\cite{yan03, yan05}, kinks may also be semi-flexible~(exhibiting higher or 
lower rigidities than the bare~DNA) or even fully flexible~\cite{yan04, wiggins}. 
For instance, the HU/DNA~complex was recently observed to be very flexible under 
specific experimental conditions~\cite{noort}. Flexible hinges were also stated to 
occur along the~DNA due to the opening of small denaturation 
bubbles~\cite{yan04, ranjith}, such as the one needed by~HU to fit in the double 
helix~\cite{lia}. Such flexibility could be taken into account~\cite{popov} in our 
model. This kind of defects could also be thermally activated, occuring at multiple 
non-specific sites along the~DNA~\cite{yan03, yan05,wiggins, popov, chakrabarti}. 
Both effects may be included in our model. Note that using effective persistence 
lengths could turn out convenient, despite these inform little about the kink 
properties~(number, location, rigidity, etc). Actually this would be equivalent to 
study DNA~stiffness loss due to sequence effects~\cite{nelson} by cutting~WLC in 
different stiff fragments, depending on the~CG or~AT content of the whole sequence 
to cyclize. The same approach may allow an approximative study of the~DNA 
polyelectrolyte nature too~\cite{zandi}. Otherwise electrostatic potential has to 
be included in WLC~energy. Twist elasticity leads to slight modifications of the 
quantum analog we used although requiring some care~\cite{bouchiat}. This is 
expected to play an important role in looping, especially when specific alignment 
of the loop extremities are required. Finally cyclization dynamics could be modeled 
using a simple two states model where~DNA is ``closed'' or ``opened'', that is 
cyclized or not. Such study relies on the~(statics) cyclization factor we computed 
in this article~\cite{jun}. Direct comparison to experimental lifetimes measures 
would be possible~\cite{finzi, lia}. 
%
%%%%%%%%%%%%%%%%%%%%%%%%%%%%%%%%%%%%%%%%%%%%%%%%%%%%%%%%%%%%%%%%%%%%% 
% 
\begin{acknowledgments}
The authors would like to thank D.~\textsc{Chatenay}, R.~\textsc{Monasson} 
and F.~\textsc{Thalmann} for useful discussions and a critical reading of 
the manuscript. 
\end{acknowledgments}
%
%%%%%%%%%%%%%%%%%%%%%%%%%%%%%%%%%%%%%%%%%%%%%%%%%%%%%%%%%%%%%%%%%%%%%%% 
%

%
%%%%%%%%%%%%%%%%%%%%%%%%%%%%%%%%%%%%%%%%%%%%%%%%%%%%%%%%%%%%%%%%%%%%%%% 
%
\end{document}